\documentclass[10pt,letterpaper]{article}
\usepackage[top=0.85in,left=1.0in,footskip=0.75in]{geometry}
\usepackage{graphicx}
\usepackage{placeins}

\usepackage{amsmath,amssymb}

\usepackage{changepage}

\usepackage[utf8x]{inputenc}

\usepackage{textcomp,marvosym}

\usepackage{cite}

\usepackage{nameref,hyperref}

\usepackage[right]{lineno}

\usepackage{microtype}
\DisableLigatures[f]{encoding = *, family = * }

\usepackage[table]{xcolor}

\usepackage{array}

\newcolumntype{+}{!{\vrule width 2pt}}

\newlength\savedwidth

\raggedright
\usepackage[aboveskip=1pt,labelfont=bf,labelsep=period,justification=raggedright,singlelinecheck=off]{caption}

\makeatletter
\renewcommand{\@biblabel}[1]{\quad#1.}
\makeatother

\date{}

\usepackage{booktabs} 
\usepackage{subfig}
\usepackage{url}
\usepackage{pbox}
\usepackage{siunitx,booktabs}
\usepackage{color,soul}      

\begin{document}
\begin{flushleft}
{\Large
  \textbf{Forex trading and Twitter: \\[1ex] Spam, bots, and reputation manipulation}
}
\newline
\\
Igor Mozeti\v{c}\textsuperscript{1},
Peter Gabrov\v{s}ek\textsuperscript{2},
Petra Kralj Novak\textsuperscript{1}
\\
\bigskip
\textsuperscript{1} Department of Knowledge Technologies, Jo\v{z}ef Stefan Institute, Ljubljana, Slovenia
\\
\textsuperscript{2} Faculty of Computer and Information Science, University of Ljubljana, Slovenia
\\
\bigskip

Correspondence: igor.mozetic@ijs.si

\end{flushleft}

\section*{Abstract}
Currency trading (Forex) is the largest world market in terms of volume.
We analyze trading and tweeting about the EUR-USD currency pair over a period of three years.
First, a large number of tweets were manually labeled, and a Twitter stance
classification model is constructed. The model then classifies all the tweets by the
trading stance signal: buy, hold, or sell (EUR vs. USD). The Twitter stance is compared
to the actual currency rates by applying the event study methodology, well-known
in financial economics. It turns out that there are large differences in Twitter stance
distribution and potential trading returns between the four groups of Twitter users:
trading robots, spammers, trading companies, and individual traders.
Additionally, we observe attempts of reputation manipulation by post festum removal of tweets
with poor predictions, and deleting/reposting of identical tweets to increase the visibility
without tainting one's Twitter timeline.
\footnote{Presented at MIS2: Misinformation and Misbehavior Mining on the Web, 
Workshop at WSDM-18, Marina Del Rey, CA, USA, February 9, 2018.}


\section{Introduction}

Foreign exchange market (Forex) is a global decentralized market for trading with currencies.
The daily trading volume exceeds 5 trillion USD, thus making it the largest market in the world.

In this paper we analyze three sources of data, over a period of three years
(from January 2014 to December 2016)~\cite{Gabrovsek2017msc}:
\begin{itemize}
\item the actual EUR-USD exchange rates,
\item financial announcements provided by the central banks (ECB and FED) and governments
that influence both currencies (called ``events''), and
\item tweets related to both currencies and their exchange.
\end{itemize}
We focus on potential missinformation spreading and manipulations on Twitter.
The main issue is: What is the ground truth?
We address this problem by moving out of the social network system and by
observing another, financial market system. Actual financial gains in the market
provide clues to potential manipulations in the social network.

We relate both systems by applying and adapting the ``event study'' methodology 
\cite{mackinlay1997event}. The currency announcements are events which are expected 
to influence the EUR-USD exchange rate. If the event signal (buy, hold, or sell) is
properly recognized then some actual financial returns can be made in the hours (or days) 
after the event. In contrast to classical event studies, we categorize events on
the basis of sentiment (properly called ``stance'') of relevant Twitter users.
In our previous work, we already analyzed the effects of Twitter stance on
stock prices (30 stocks from the Dow Jones index) \cite{Ranco2015effects,Gabrovsek2017twitter}.
We showed that the peaks of Twitter activity and their polarity are
significantly correlated with stock returns.
In this paper, we show that, for certain classes of Twitter users, returns after the events
are statistically significant (albeit small). And we can also identify
differences in returns after the potential manipulations of Twitter feed.

The paper is organized as follows.
In section \ref{sec:stance} we specify how the Forex tweets were collected,
a subset manually annotated, and a stance classification model constructed.
Section \ref{sec:user_groups} provides simple rules to identify different
classes of Twitter users (such as trading robots, spammers, and actual traders).
We show that there are large differences in Twitter stance between these users.
Section \ref{sec:event_study} describes the event study methodology in some
detail, as needed to understand the subsequent results. We show significant
differences in cumulative abnormal returns between the different user groups.
In section \ref{sec:reputation} we address potential manipulations of the user
Twitter feed with a tentative goal to improve her/his reputation and visibility.
We focus on the tweets that were deleted after we originally collected them, 
and analyze different reasons for this post festum deletions. 
We conclude with the ideas for further work and enhancements of the preliminary,
but promising, results presented so far.

\section{Twitter stance model}
\label{sec:stance}

Tweets related to Forex, specifically to EUR and USD, were acquired through the
Twitter search API with the following query: ``EURUSD'', ``USDEUR'', ``EUR'', or ``USD''.
In the period of three years (January 2014 to December 2016) almost 15 million
tweets were collected. A subset of them (44,000 tweets) was manually labeled
by knowledgeable students of finance. The label captures the leaning or stance of
the Twitter user with respect to the anticipated move of one currency w.r.t. the other.
The stance is represented by three values: buy (EUR vs. USD), hold, or sell.
The tweets were collected, labeled and provided to us by the Sowa Labs company
(\url{http://www.sowalabs.com}).

The labeled tweets were generalized into a Twitter stance model.
For supervised learning, variants of SVM \cite{Vapnik1995} are often used,
because they are well suited for large scale text categorization, are robust, and perform well.
For Forex tweets, we constructed a two plane SVM classifier \cite{Mozetic2016multilingual,Mozetic2018howto}.
The two plane SVM assumes the ordering of stance values and implements ordinal classification. 
It consists of two SVM classifiers: One classifier is trained to separate the `buy' tweets from the
`hold-or-sell' tweets; the other separates the `sell' tweets from the `buy-or-hold' tweets.
The result is a classifier with two hyperplanes that partitions the 
vector space into three subspaces: buy, hold, or sell. During classification,
the distances from both hyperplanes determine the predicted stance value.

The stance classifier was evaluated by 10-fold blocked cross-validation.
Since tweets are time-ordered, they should not be randomly selected into individual folds,
but retained in blocks of consecutive tweets \cite{Cerqueira2017comparative}.
The results of performance evaluation are in Table \ref{table:label_model_validation}.
Note that the F$_1$ measure considers just the `buy' and `sell' classes, as is common
in the three-valued sentiment classification evaluations \cite{Mozetic2016multilingual}.

\begin{table}[ht!]
    \centering
    \begin{tabular}{|r|c|}
        \hline
        \textbf{Measure} & \textbf{Value}    \\ \hline
        Accuracy         & $0.811 \pm 0.014$ \\
        F$_1(buy, sell)$        & $0.810 \pm 0.014$ \\ \hline
    \end{tabular}
    \caption{Evaluation results of the Twitter stance model.}
    \label{table:label_model_validation}
\end{table}

\section{Twitter user groups}
\label{sec:user_groups}

Different types of Twitter users have very different intentions regarding
their impact and message they want to spread. In recent years, specially
automated robots became increasingly influential. To properly estimate
the relation between the Forex market and tweetosphere, it is important
to focus on relevant Twitter users, i.e., Forex trading companies and
individual traders.

In related work, it was already shown that bots exercise a profound impact 
on content popularity and activity on Twitter. For example, Gilani et al.
\cite{gilani2017bots} implemented a simple bot detection mechanism 
based on click frequency and user agent strings. To classify users into
three categories (organizations, journalists/media bloggers, and 
individuals), De Choudhury et al. \cite{de2012unfolding} trained an
automatic classifier. An alternative approach is to detect communities
in a retweet network, e.g., \cite{cherepnalkoski2016retweet,Sluban2015sentiment}.

It turns out that it is easy to identify Forex trading robots. 
Their tweets ($t(bots)$) all start with one of the eighth patterns (such as ``Closed Buy'',
``Sell stop'', ...). The Forex Twitter users can then be classified into one
of the four groups by the following simple rules:
\begin{itemize}
\item Trading robots:\\ $t(bot)_{rate} > 0.75$
\item Spam/scam/advertisements:\\ $tweets > 1000$ \& $retweeted_{ratio} < 0.01$
\item Trading companies:\\ $days_{active} > 30$ \& $t_{rate} > 0.5$ \& $retweeted_{ratio} > 0.25$
\item Individual traders:\\ $days_{active} > 30$ \& $retweeted_{ratio} > 0.05$
\end{itemize}
where $t_{rate} = tweets/days_{active}$ indicates the daily activity of the user, and
$retweeted_{ratio} = retweeted/tweets$ is the proportion of the user tweets that were retweeted
by others.

\begin{figure}[ht!]
\centering
\includegraphics[width=15cm]{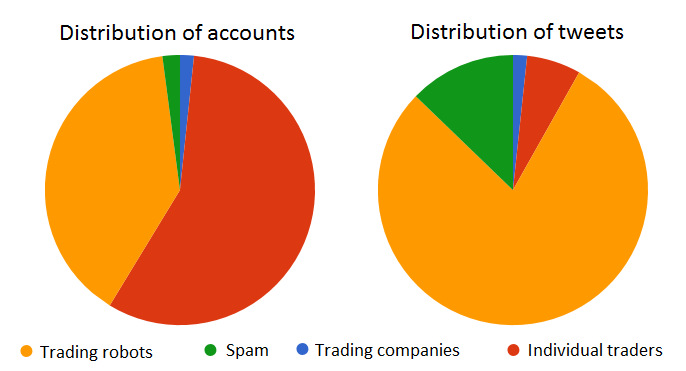}
\caption{Proportions of Twitter accounts and tweets for different user groups.}
\label{fig:user_groups}
\end{figure}

Figure \ref{fig:user_groups} shows the proportions of different Twitter user groups and their tweets
in our dataset. We can see that more than half of the users are individuals, but that the trading
robots produce by far the largest fraction of Forex tweets.

\begin{figure*}[ht!]
\centering
\includegraphics[width=15cm]{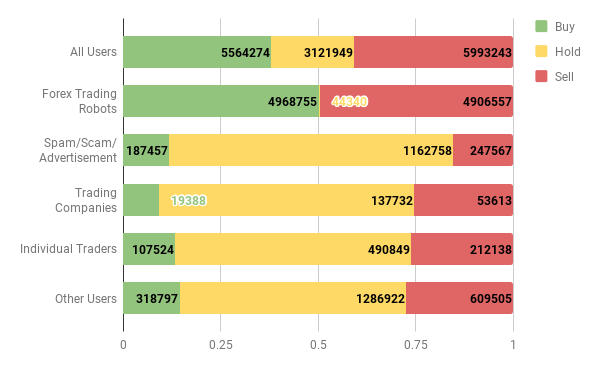}
\caption{Twitter stance distribution of different user groups (bars show the proportion of tweets).
Trading robots produce almost exclusively polarized tweets while spammers are predominantly neutral.
}
\label{fig:sentiment_distributions}
\end{figure*}

There are also considerable differences in the stance between different user groups.
Figure \ref{fig:sentiment_distributions} shows that trading robots produce almost
exclusively polarized tweets (no `hold' tweets). On the other hand, spammers (without robots)
are predominantly neutral (relatively few `buy' or 'sell' tweets). The groups we focus on,
trading companies and individuals, are more opinionated than spammers.
It is interesting that in their tweets the `sell' signal is prevailing, probably due 
to the downward trend of EUR vs. USD in the last three years.

\newpage
\section{Event study}
\label{sec:event_study}

An event study captures the impact of external events on the market returns.
External events that we consider here are the currency related announcements
by the central banks (FED and ECB) and governments (around 750 in the three years).
In an event study, Cumulative Abnormal Return (CAR) is defined as a measure 
of return which exceed the overall market return. Specifically:

\begin{itemize} 
\item \textbf{Market model} corresponds to the overall market movement
before the event. In our case, we use a linear regression of 30 days 
currency ratios prior to the event. The market model price is then
subtracted from the actual currency price (at one minute resolution)
to get the abnormal price ($pab$):
$$
pab_{i} = p_i - k * i
$$ 
where $p_i$ is actual price at time $i$ after the event.

\item \textbf{Abnormal return} is computed as a relative (abnormal) price change:
$$
rab_{i} = \frac{pab_{i + 1} - pab_{i}}{pab_{i}}
$$

\item \textbf{Cumulative abnormal return (CAR)} measures aggregated 
returns over longer periods of time $i$:
$$
CAR = \sum_{i = 0}^n rab_{i}
$$
\end{itemize}

The other essential component of an event study is determining the 
\textbf{type of event} in terms of its expected impact on the price. 
In stock market, typically Earnings Announcements are studied. 
If an announcement exceeds prior expectations of analysts, it is classified as positive, 
and stock prices are expected to rise. An event study combines announcements about
several stocks, over longer period of time, and computes the average
CARs in the days or hours after the announcements.

In our case, we do not consider expectation of the analysts, but instead use
the stance of the Forex Twitter users regarding the EUR vs. USD exchange rate.
We consider all tweets in one hour after the announcement, and aggregate
their stance to categorize the event. Then we compute the CARs for up to one day
after the event, at one minute resolution. If Twitter stance correctly predicts
the exchange rate movement then there should be some tangible returns
(CARs) in the hours after the event.

\begin{figure*}[ht!]
\centering
\includegraphics[width=\textwidth]{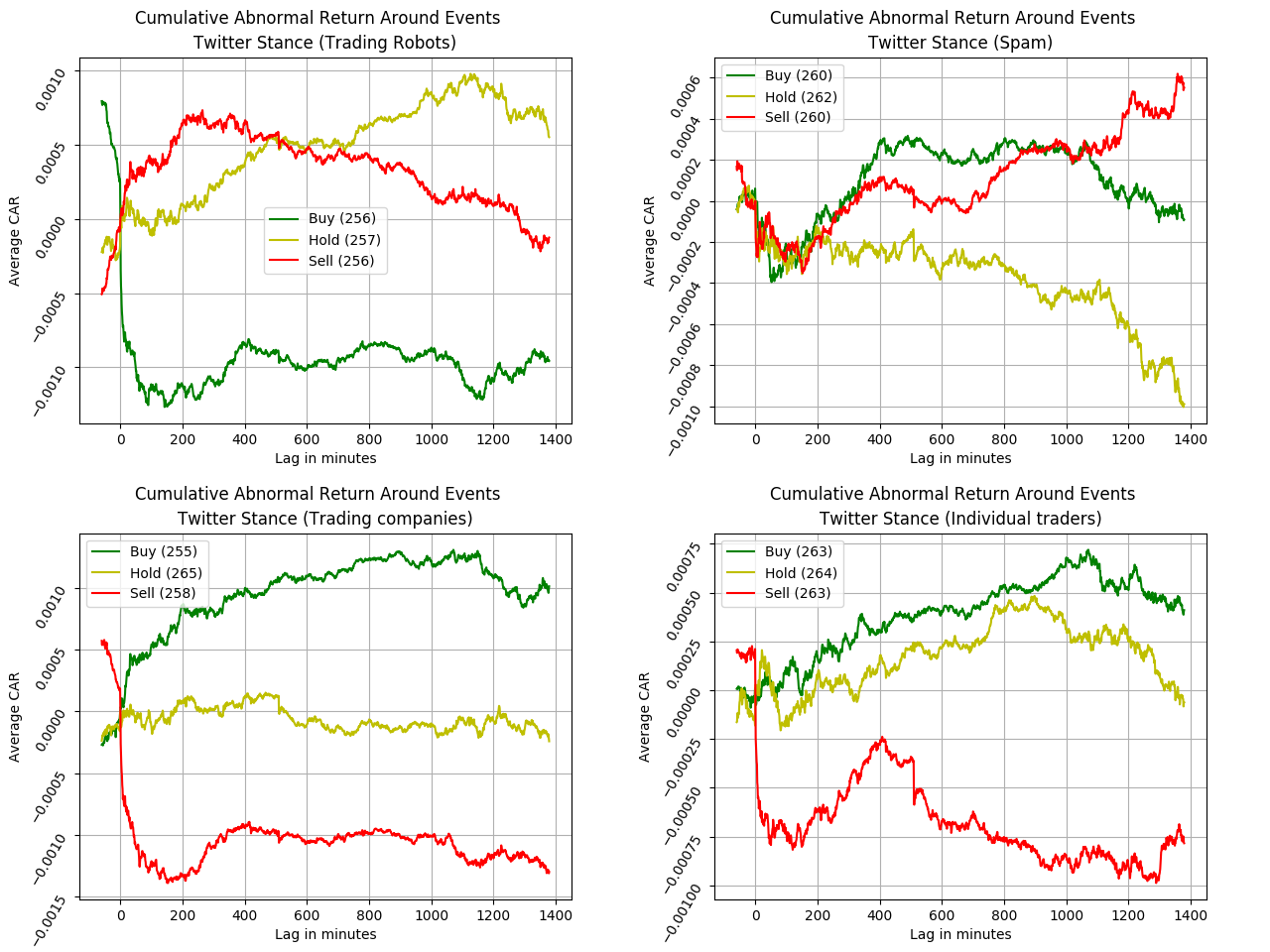}
\caption{Cumulative abnormal returns (CARs) for different user groups. 
The events are classified as `buy', `hold', or `sell' 
according to the cumulative Twitter stance in one hour after the event.
The event is announced at lag = 0.
CARs are computed at one minute resolution, for up to one day (1440 minutes)
after the event.}
\label{fig:car_joined_user_groups}
\end{figure*}

Figure \ref{fig:car_joined_user_groups} shows returns, aggregated over all 750
events, for different Twitter user groups. The expected result is visible for
trading companies (bottom-left chart). For `buy' events  (we buy EUR at time 0)
CARs are positive (return is around 0.1\%, small but significant), for `sell' events 
(we sell EUR at time 0) CARs are negative , and for `hold' events (no transaction) 
CARs are around zero. Similar results are obtained for individual traders
(bottom-right chart), but the separation of events is not as clear as for
trading companies.

On the other hand, trading robots and spam users
(top two charts in Figure \ref{fig:car_joined_user_groups}) show no useful
correlation between the Twitter stance and CARs. As a consequence, we conclude
that it is important to properly identify them and eliminate their tweets from
any trading strategy based on Twitter.

\newpage
\section{Reputation manipulation}
\label{sec:reputation}

Here we focus on another aspect of Twitter misuse for potential manipulation:
post festum deletion of tweets by the Twitter user. What are the reasons for
users to delete their tweets? 
Previous research addressed prediction of malicious or deleted tweets
\cite{martinez2013detecting, almuhimedi2013tweets, petrovic2013wish},
and identification of deleted and suspicious accounts \cite{volkova2017identifying}. 
On one hand, some authors show that typos and rephrasing are among the major causes for
deleting tweets \cite{almuhimedi2013tweets}. On the other hand, other authors found
that in deleted tweets, a significantly higher fraction of the vocabulary consists of 
swear words, and markers that indicate anger, anxiety, and sadness \cite{bhattacharya2016characterizing}.

\begin{figure}[ht!]
\centering
\includegraphics[width=15cm]{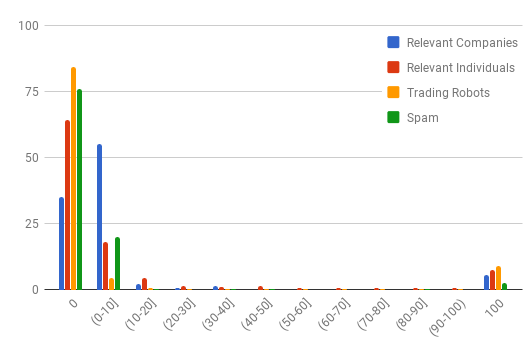}
\caption{Fractions of tweets deleted for different user groups.}
\label{fig:deleted_tweets}
\end{figure}

We verified which of the tweets that were collected during the three years in near
real time, still exist. It turns out that in our dataset, 4.7\% (689,658) posts were post festum
deleted by the users. Different user groups exhibit different patterns of deletion.
A histogram in Figure~\ref{fig:deleted_tweets} shows fractions of tweets deleted by different user groups.
The majority of users do not delete their own tweets at all (peak at 0). At the other extreme (100),
there is about 5\% of the users who deleted their accounts and all their tweets.
But the really interesting are the trading companies, where only one third of them
does not delete tweets, and more than half of them delete up to 10\% of their tweets.

We focus on the deleted tweets by {\it trading companies} and {\it individual traders} 
and search for signs of reputation manipulations.
A breakdown of deleted tweets for both groups in terms of different stances is in Table~\ref{table:deleted_sentiment}.

\begin{table}[ht!]
    \centering
    \begin{tabular}{|l|rrr|}
    \hline
    User group  & Buy           & Hold           & Sell           \\ \hline
    Trading     &               &                &                \\
    companies   & 453 (2.3\%)   & 3,285 (2.4\%)  & 1,297 (2.4\%)  \\ \hline
    Individual  &               &                &                \\
    traders     & 4,438 (4.1\%) & 35,915 (7.3\%) & 11,572 (5.5\%) \\ \hline
    \end{tabular}
    \caption{The number of deleted tweets of different stance.}
    \label{table:deleted_sentiment}
\end{table}

\newpage
\subsection{Deleting tweets to increase CARs}

One reason for companies and individuals to delete their tweets might be to 
create an image of their capabilities to predict the market.
For example, one can post two contradictory tweets at the same time:
EUR will go up, and EUR will go down. After the market shows the actual EUR move,
the incorrect prediction is deleted, and the user's timeline shows his
forecasting insight.

\begin{figure*}[ht!]
    \centering
    \subfloat{{\includegraphics[width=10cm]{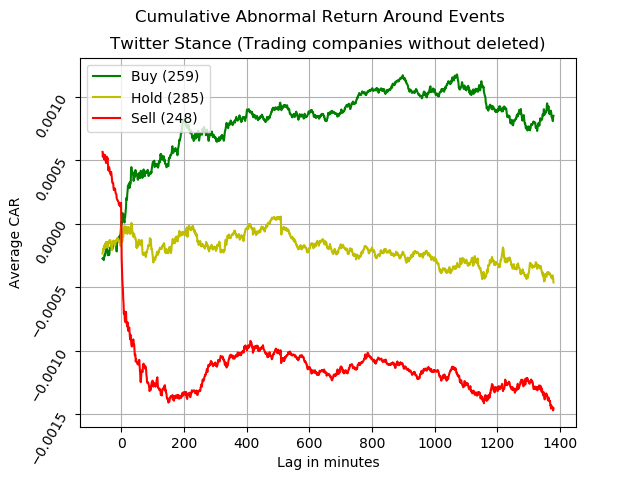}}}%
    \qquad
    \subfloat{{\includegraphics[width=10cm]{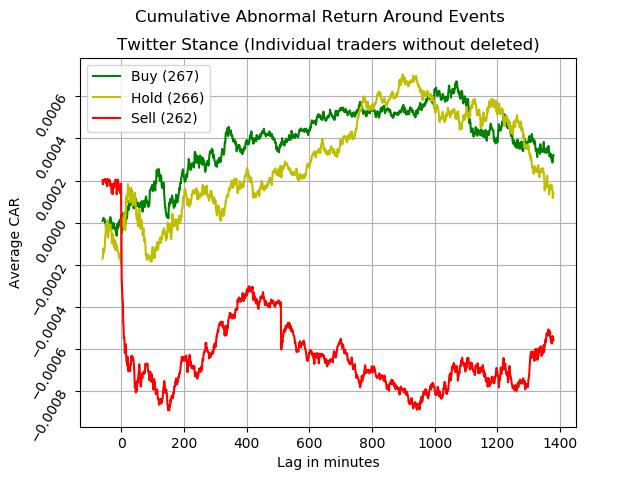}}}%
    \caption{Cumulative abnormal returns (CARs) for trading companies and individual traders, 
after removing the tweets that were post festum deleted by the user.}
    \label{fig:car_comp_individuals}
\end{figure*}

We compare the results of the event study before and after the tweets were deleted.
Figure \ref{fig:car_comp_individuals} shows CARs for trading companies and individual
traders after removing their deleted tweets. At this point, we can report only
negative results, i.e., there is no increase of CARs, and the `hold' events are
further away from the zero line than in Figure \ref{fig:car_joined_user_groups}.

\subsection{Analyzing trading companies}

\begin{figure}[ht!]
\centering
\includegraphics[width=12cm]{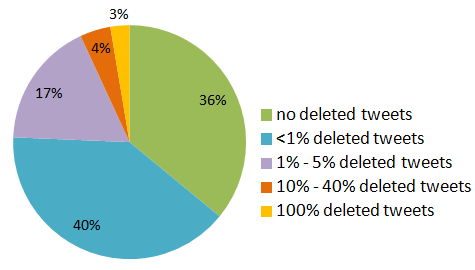}
\caption{The fractions of deleted tweets (altogether 3,741 tweets) for the 189 trading companies.}
\label{fig:proportion_deleted_tweets_relevant_comp}
\end{figure}

We analyze deleted tweets of 189 (out of 195) Twitter users categorized as {\it trading companies} that have 
active Twitter accounts (by deleting an account, all the tweets from that account are also deleted).
The 189 companies deleted 3,741 tweets. Among them, 
four deleted all Forex related tweets from their profile while the accounts are still active, 
8 users deleted between 10\% and 40\% of their tweets, 
33 users deleted between 1\% and 5\% of their tweets, 
and only 68 did not delete any tweets. The deleting behaviour of trading companies is shown 
in Figure~\ref{fig:proportion_deleted_tweets_relevant_comp}.
Note that the majority (76\% of the trading companies) deleted less than 1\% of their tweets. 
Note also that there are no trading companies that delete between 5 and 10\% of their tweets.
We analyze the deleted tweets and focus on criteria that might indicate reputation manipulation.

\begin{figure}[ht!]
\centering
\includegraphics[width=15cm]{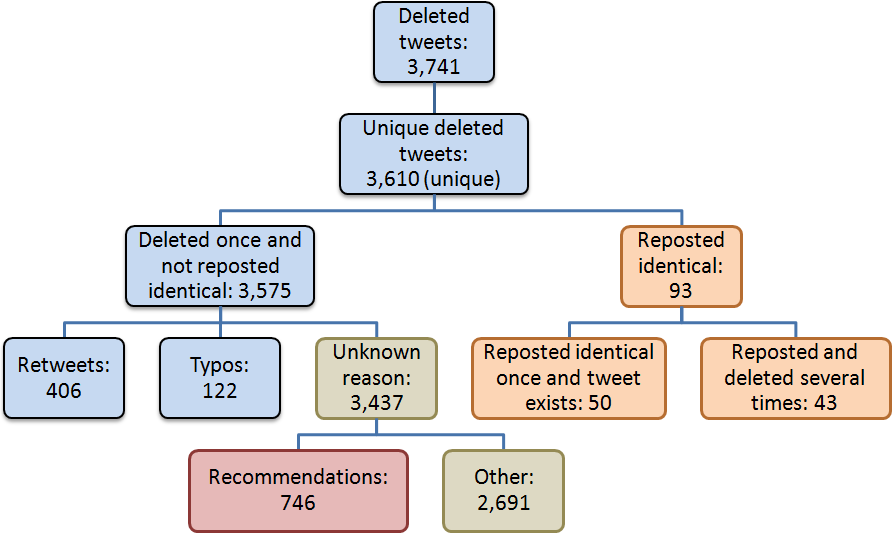}
\caption{A breakdown of deleted tweets by trading companies.}
\label{fig:deleted_tweets_relevant_comp}
\end{figure}

Out of the 3,741 deleted tweets, 3,611 are unique (same author and identical text) 
while 130 tweets are deleted more than once. An extreme case is a tweet (advertising easy and safe profit) 
which is deleted 46 times (same author and identical text). The deleting and reposting of identical tweets 
is one form of increasing visibility without tainting the author's Twitter timeline. 
A tweet that is deleted and posted again appears several times in the user's followers feed
while it appears just once in the authors timeline. 
This can be therefore considered a kind of reputation manipulation.
Out of the 93 tweets that were deleted and reposted, 50 were deleted and reposted once while the rest 
were deleted and reposted several times.
The 746 `recommendation' tweets that were deleted afterward point to a potential 
reputation manipulation by deleting the bad recommendations.
The breakdown of deleted tweets is shown in Figure~\ref{fig:deleted_tweets_relevant_comp}.

One of the major reasons to delete tweets are typos and rephrasing \cite{almuhimedi2013tweets}.
In these cases, a very similar tweet to the deleted tweet is posted again. 
We check for each of the 3,575 tweets that were deleted once and not reposted, if they were deleted due to a typo. 
We define typo as a reason of tweet deletion if the tweet is:
\begin{itemize}
\item posted by the same author,
\item within the three next tweets after the deleted one,
\item with a very similar text ($1 <$ Levenshtein distance $< 4$),
\item and the difference is not in the URLs present in the tweet.
\end{itemize}
We found that 122 deleted tweets were reposted with changes so small that indicate typos.

Another category of deleted tweets are retweets. 
If retweets are deleted, it is usually because the original tweets were deleted.
In our dataset, 406 retweets are deleted.

We check the remaining 3,437 tweets for the use of vocabulary specific for trading:
long, short, bear, bull,  bearish, bullish, resistance, support, buy, sell, close. 
We identify 746 tweets that are recommendations for trading (manually confirmed).
This is another kind of possible reputation manipulation: a tweet with recommendation is posted and afterwards, 
if the recommendation turns out to be spurious, the tweet is deleted.
The author's Twitter timeline then falsely appears as if following his recommendations would yield profit. 

We inspect a specific Twitter account from the category {\it trading companies}
that posted more than 500 tweets and deleted between 10\% and 40\% of them.
The identity of the account cannot be revealed due to the privacy issues.
The tweets deleted fall into the following categories:
\begin{itemize}
\item Reposts: 91, 60 of them are advertisements (e.g., subscribe for analysis),
\item Links (to recommendations): 17,
\item Recommendations: 11,
\item Retweet: 1 (if the original tweet is deleted, retweets are also deleted).
\end{itemize}
    

We manually checked each of the 11 recommendations that were deleted. 
In all the cases, the recommendations turned out to be bad, i.e., an investor would loose money.
An (anonymized) example of a bad recommendation post is the following:

"@user\_mention while daily candle is above 1.xyz we are bullish on \$EURUSD."

while in the actual Forex market, EUR went down.

This user used both types of reputation manipulation: deleting poor recommendations, and deleting/reposting of identical tweets to increase their visibility.
The percentage of deleted poor predictions is small compared to all the deleted tweets and compared to all the posted tweets. 
We speculate that the manipulation by tweet deletion needs to be subtle to go unnoticed by the users' followers. 
However, even a subtle reputation burst in a domain as competitive as Forex trading can bring major benefits to the deceptive user.

\section{Conclusions}

This is an initial study of potential misuses of Twitter to influence
the public interested in Forex trading.
We identify different types of Twitter accounts that are posting tweets
related to the EUR-USD currency exchange. We show that there are considerable
differences between them in terms of Twitter stance distribution and CARs.
If we eliminate trading robots and spam, we find significant
correlations between the Twitter stance and CARs (the returns are small,
but the Forex market has very low trading costs).
The remaining posts come from the Forex trading companies and individual traders.
We further analyze the reasons for post festum deleting of tweets.
Some reasons are harmless (such as correcting typos), but some
show indications of reputation busting.
We consider this a promising direction for further, more in-depth
analysis.

\subsection*{Acknowledgements}

The authors acknowledge financial support from the H2020 FET project DOLFINS (grant no. 640772), 
and the Slovenian Research Agency (research core funding no. P2-103).

\end{document}